\documentclass[aps,prsty,floatfix,twocolumn]{revtex4}
\usepackage{graphics}
\usepackage{dcolumn}
\usepackage{epsfig}

\newcommand{\simlt}
      {\ifmmode       { \raisebox{-.8em}{$<$}\atop\sim}
         \else        {$\raisebox{-.8em}{$<$}\atop\sim$}
      \fi}

\newcommand{\bk}{{\bf k }}

\newcommand{\bq}{{\bf q }}
\newcommand{\br}{{\bf r }}

\newcommand{\bG}{{\bf G}}

\newcommand{\bK}{{\bf K}}

\newcommand{\EP}{{\it e}-ph}

\newcommand{\EE}{{\it e}-{\it e}}

\begin{document}

\bibliographystyle{prsty}

\title
{First-Principles Study of Electron Linewidths in Graphene}
\author{Cheol-Hwan Park$^{1,2}$}
\author{Feliciano Giustino$^{1,2,3}$}
\author{Catalin D. Spataru$^4$}
\author{Marvin L. Cohen$^{1,2}$}
\author{Steven G. Louie$^{1,2}$}
\email{sglouie@berkeley.edu}
\affiliation{$^1$Department of Physics, University of California at Berkeley,
Berkeley, California 94720 USA\\
$^2$Materials Sciences Division, Lawrence Berkeley National
Laboratory, Berkeley, California 94720 USA\\
$^3$Department of Materials, University of Oxford, Oxford, OX1 3PH,
United Kingdom\\
$^4$Sandia National Laboratories, Livermore, California 94551 USA}

\date{\today}

\begin{abstract}
We present first-principles calculations of the linewidths of
low-energy quasiparticles in {\it n}-doped graphene
arising from both
the electron-electron and the electron-phonon
interactions. The contribution to the
electron linewidth arising from the electron-electron interactions
vary significantly with wavevector at fixed energy;
in contrast, the electron-phonon contribution is virtually
wavevector-independent.
These two contributions are comparable in magnitude
at a binding energy of $\sim0.2$~eV, corresponding to the optical phonon energy.
The calculated linewidths, with both electron-electron and
electron-phonon interactions included, explain
to a large extent the linewidths seen in
recent photoemission experiments.
\end{abstract}
\maketitle

Graphene~\cite{novoselov:2005Nat_Graphene_QHE,
zhang:2005Nat_Graphene_QHE,berger:2006Sci_Graphene_Epitaxial},
a single layer of carbon atoms in a hexagonal honeycomb structure,
is a unique system whose carrier dynamics can be described by a massless Dirac
equation~\cite{wallace:1947PR_BandGraphite}.
Within the quasiparticle picture, carriers in graphene
exhibit a linear energy dispersion relation and chiral behavior
resulting in a half-integer
quantum Hall effect~\cite{novoselov:2005Nat_Graphene_QHE,
zhang:2005Nat_Graphene_QHE}, absence of
backscattering~\cite{ando:1998JPSJ_NT_Backscattering,mceuen:1999PRL_NT_Backscattering},
Klein tunneling~\cite{katsnelson:2006NatPhys_Graphene_Klein}, and
novel phenomena such as electron supercollimation in
superlattices~\cite{park:2008NatPhys_GSL,park:2008NL_Supercollimation,park:126804}.

Graphene is considered a promising candidate for
electronic and spintronic devices~\cite{geim:2007NatMat_Graphene_Review}.
For these applications it is important to understand the effects of
many-body interactions
on carrier dynamics.
In particular, the scattering rate of charge carriers, manifested in their
linewidths, affects
the transport properties of actual devices.

The scattering of charge carriers in solids
can arise from several different
mechanisms, among which electron-hole pair generation, electron-plasmon
interaction, and electron-phonon (\EP) interaction are generally important.
Scattering by impurities, defects and interactions
with the substrate also
affects the carrier dynamics.
The contribution to the electron linewidths arising from the \EP\
interaction has been studied with first-principles
calculations~\cite{park:2007PRL_Graphene_ElPh,park:2008PRB_Graphene_ElPh}
and through the use of analytical and numerical
calculations based on the massless Dirac equation~\cite{calandra:205411,tse:236802}.
The linewidth contribution originating from electron-electron (\EE) interactions,
which includes both the electron-hole pair generation process and the electron-plasmon interaction,
has only been studied within the massless Dirac equation
formalism~\cite{polini:081411,hwang:081412,hwang:115434}.

A recent angle-resolved photoemission experiment on {\it n}-doped graphene
epitaxially grown on silicon carbide (SiC)~\cite{bostwick:2007NatPhys}
has stimulated experimental~\cite{ohta:2007PRL_Graphene_ARPES,
mcchesney:2007arxiv_Graphene_Anisotropy,zhou:2007NatMat}
and theoretical~\cite{polini:081411,hwang:081412,hwang:115434,park:2007PRL_Graphene_ElPh,
calandra:205411,tse:236802} studies on this topic.
In Ref.~\onlinecite{bostwick:2007NatPhys}, the width of the momentum distribution curve (MDC)
from photoemission data is presented.
The MDC of the graphene photoemission spectra
is observed to resemble a simple Lorentzian whose width may be interpreted to be directly
proportional to the scattering rate~\cite{bostwick:2007NatPhys}.

We draw the attention to the well-known controversy in
the different interpretations of the angle-resolved
photoemission spectra of graphene.
It is claimed in Ref.~\onlinecite{bostwick:2007NatPhys}
that the spectral features can entirely be understood from
many-body effects, including both \EE\ and \EP\ interactions,
in graphene. On the other hand, in Ref.~\onlinecite{zhou:2007NatMat},
it is argued that many of those features are dominated by
an energy gap of 0.2$\sim$0.3~eV, which opens up at the Dirac
point energy ($E_{\rm D}$) because of interactions between graphene and
the reconstructed surface of SiC.
This important problem in understanding the quasiparticle
spectra of graphene (which also have implications
in graphene-based electronics applications) has led to numerous
additional experiments directly or indirectly addressing
this discrepancy~\cite{ohta:2008NJP,bostwick:2007NJP,
rotenberg:2008NatMat_comment,zhou:2008NatMat_re,zhou:2008PhysicaE,zhou:086402}.
On the theoretical side, several density functional theory
calculations on the effect of substrates without considering
many-body effects, along the line of Ref.~\onlinecite{zhou:2007NatMat},
have been performed~\cite{mattausch:2007prl,varchon:126805,kim:176802}.
On the other hand, first-principles calculations
on the effects of both \EE\ and \EP\ interactions,
along the line of Ref.~\onlinecite{bostwick:2007NatPhys},
have been lacking up to now.

In this paper, to fill in this missing part, we present
{\it ab initio} calculations of
the electron linewidth in {\it n}-doped graphene arising
from \EE\ interactions employing the {\it GW}
approximation~\cite{hybertsen:1986PRB_GW,spataru:246405,note:realpart}.
In addition, we calculate the electron linewidth
originating from the \EP\ interaction following
the method in Refs.~\onlinecite{park:2007PRL_Graphene_ElPh,park:bilayer}
and~\onlinecite{giustino:2007PRB_Migdal}.
Combining both contributions,
we provide a comprehensive view of the scattering rate
originating from many-body effects.
Our calculation indicates that the linewidth arising from \EE\
interactions is highly anisotropic.
This is in contrast to the insensitivity
to wavevector of the phonon-induced
electron linewidth shown in Ref.~\onlinecite{park:2008PRB_Graphene_ElPh}.
The calculated linewidth arising from \EE\ interaction
becomes comparable to that arising from \EP\ interaction at a binding energy
of $\sim0.2$~eV (i.\,e.\,, the optical phonon energy).
The combination of the two contributions accounts for most
of the measured linewidth over the 0~eV~$\sim$~2.5~eV binding energy range.

The electronic eigenstates $\left|n\bk\right>$ of graphene
are obtained with {\it ab initio}
pseudopotential density-functional calculations~\cite{cohenlouie} in the local density
approximation (LDA)~\cite{ceperley:1980PRL_pseudopotential,perdew:1981PRB_exchcorr}
in a supercell geometry. Electronic wavefunctions
in a $72\times72\times1$ {\bf k}-point grid
are expanded in a plane-waves basis~\cite{ihm:1979JPC_PW}
with a kinetic energy cutoff of 60~Ry. The  core-valence interaction
is treated by means of norm-conserving
pseudopotentials~\cite{troullier:1991PRB_pseudopotential}.
Graphene layers between adjacent supercells
are separated by 8.0~\AA\
and the Coulomb interaction is truncated to prevent
spurious interaction between periodic replicas~\cite{ismail-beigi:233103}.
Increasing the interlayer distance to
16.0~\AA\ makes virtually no difference in the calculated electron self-energy.
Doped graphene is modeled by an extra electron density
with a neutralizing background.

We calculate the imaginary part of the
electron self-energy induced by the \EE\ interaction
within the {\it GW} approximation~\cite{hybertsen:1986PRB_GW,spataru:246405}.
The frequency dependent dielectric matrices $\epsilon_{\bG,\bG'}(\bq,\omega)$
are calculated within the random phase approximation using the LDA wavefunctions
on a regular grid of $\omega$
with spacing $\Delta\omega=$0.125~eV~\cite{benedict:085116},
and a linear interpolation is performed to obtain
the dielectric matrices for energies in between the grid points.
In the calculation of the polarizability, for numerical convergence,
an imaginary component of magnitude $\Delta\omega$ of 0.125~eV as above
is introduced in the energy denominator.
Convergence tests showed that the dimension of $\epsilon_{\bG,\bG'}$
may be truncated at a kinetic energy cutoff of $\hbar^2G^2/2m=$12~Ry.
To take into account the screening of the SiC substrate,
we have renormalized the bare Coulomb interaction by an effective background
dielectric constant of 
$\varepsilon_{\rm b}=(1+\varepsilon_{\rm SiC})/2=3.8$~\cite{bostwick:2007NatPhys,
polini:081411,hwang:081412,hwang:115434}, where $\varepsilon_{\rm SiC}(=6.6)$
is the optical dielectric constant of
SiC~\cite{logothetidis:1996jap,note:epsilon}.

\begin{figure}
\includegraphics[width=1.0\columnwidth]{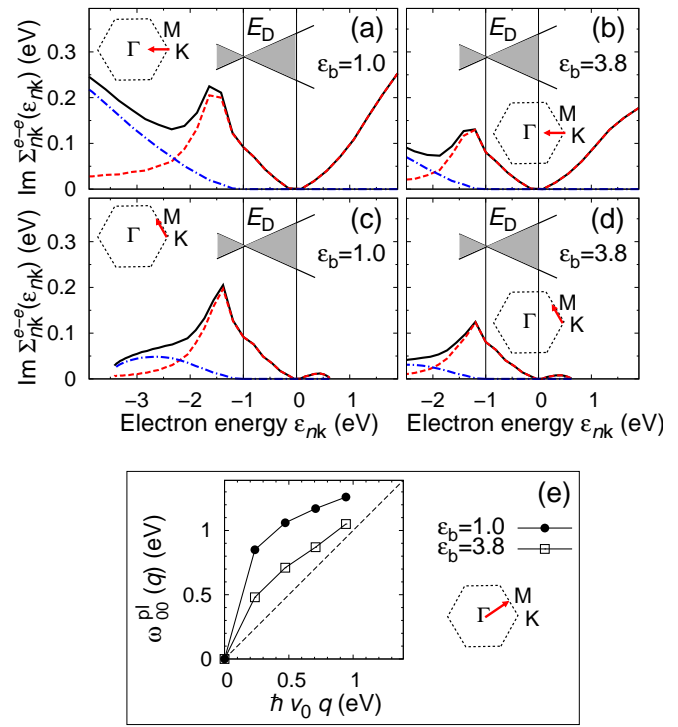}
\caption{(color online).
  (a)-(d): Calculated imaginary part of the
  electron self-energy arising from the \EE\ interaction,
  ${\rm Im}\,\Sigma^{\EE}_{n\bk}(\varepsilon_{n\bk})$, versus
  the LDA energy $\varepsilon_{n\bk}$ (solid black lines) in
  {\it n}-doped graphene. The Dirac point energy $E_{\rm D}$ is
  1.0~eV below the Fermi level.
  The contributions to ${\rm Im}\,\Sigma^{\EE}_{n\bk}(\varepsilon_{n\bk})$
  coming from electronic transitions to the upper linear bands and to the
  lower linear bands are shown as dashed red lines and
  dash-dotted blue lines, respectively.
  The self-energy is evaluated along the reciprocal space segments
  shown in the insets.
  (a) and (c) are results for suspended graphene with a background
  dielectric constant of $\varepsilon_{\rm b}=1.0$,
  whereas (b) and (d) are results for graphene with a background dielectric
  constant of
  $\varepsilon_{\rm b}=(1+\varepsilon_{\rm SiC})/2=3.8$.
  The Fermi level and $E_{\rm D}$ are indicated by vertical lines.
  (e): Calculated plasmon energy dispersion relation
  $\omega^{\rm pl}_{00}(\bq)$, given by
  $\epsilon_{\bG=0,\bG'=0}[\bq,\omega^{\rm pl}_{00}(\bq)]=0$,
  versus $\hbar v_0|\bq|$ along the $\Gamma$M direction.
  The solid lines are guides
  to the eye and the dashed line corresponds to $\omega(q)=\hbar v_0 q$.}
\label{Fig1}
\end{figure}

Figure~\ref{Fig1} shows the calculated imaginary part
${\rm Im}\,\Sigma^{\EE}_{n\bk}(\varepsilon_{n\bk})
=\left<n\bk\right|{\rm Im}\,\Sigma^{\EE}(\br,\br',\varepsilon_{n\bk})\left|n\bk\right>$
of the electron self-energy arising from the \EE\ interaction
with $\omega$ set at the LDA eigenvalue $\varepsilon_{n\bk}$.
The Fermi level $E_{\rm F}~(=0)$ is taken to be
1~eV above $E_{\rm D}$.
In Fig.~\ref{Fig1}(a), ${\rm Im}\,\Sigma^{\EE}_{n\bk}(\varepsilon_{n\bk})$
for graphene without including substrate screening, appropriate for
suspended graphene~\cite{bolotin:SSC2008,du:nnano2008},
is plotted along the K$\Gamma$ direction.
Generally, the self-energy increases with increasing $|\varepsilon_{n\bk}|$
as measured from $E_{\rm F}$.
A notable feature is the peak around $\varepsilon_{n\bk}=-1.5$~eV.
To find the origin of this peak, we have decomposed the total
electron self-energy into the contributions arising from transitions into
the upper linear bands (above $E_{\rm D}$) and the lower linear bands
(below $E_{\rm D}$).
The former involves electron-plasmon interaction~\cite{polini:081411}.
The peak structure comes from scattering processes of electrons
into the upper linear bands,
whereas those scattering processes into the lower linear bands result in a monotonic
increase in the electron linewidth.
When the background dielectric
constant $\varepsilon_{\rm b}$ is
changed from 1 to 3.8 [Fig.~\ref{Fig1}(b)],
the position of this peak shifts toward lower-binding
energy by $\sim0.3$~eV,
reflecting a decrease of the plasmon energy
in graphene [Fig.~\ref{Fig1}(e)]~\cite{wunsch:2006,hwang:205418}.
The height of the peak is further suppressed.
At low energy ($|\varepsilon_{n\bk}|<1.0$~eV), the imaginary part of
the self-energy is however not sensitive to the choice of $\varepsilon_{\rm b}$.

\begin{figure}
\includegraphics[width=1.0\columnwidth]{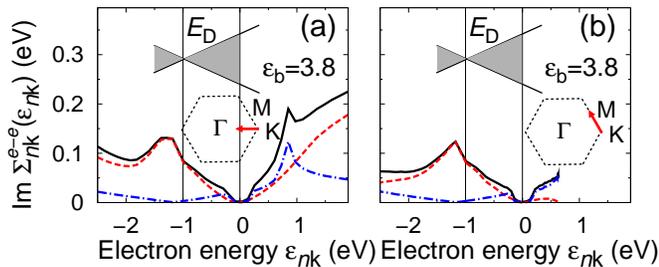}
\caption{(color online). Calculated
  Im~$\Sigma_{n\bk}(\varepsilon_{n\bk})$ versus the LDA energy eigenvalue
  $\varepsilon_{n\bk}$ in
  {\it n}-doped graphene ($E_{\rm D}=-1.0$~eV) on a model substrate ($\varepsilon_{\rm b}=3.8$).
  The total self-energy, the self-energy arising from the \EE\ interaction,
  and that arising from the \EP\ interaction are shown in
  solid black, dashed red and dash-dotted blue lines, respectively.
  The self-energy is evaluated along the reciprocal space segments
  shown in the insets.}
\label{Fig2}
\end{figure}

Comparing Figs.~\ref{Fig1}(a) and~\ref{Fig1}(b) with Figs.~\ref{Fig1}(c)
and~\ref{Fig1}(d) shows that
the electron self-energy arising from
the \EE\ interaction calculated along the KM direction
is very different from
that along the K$\Gamma$ direction.
Below $-1.5$~eV, ${\rm Im}\,\Sigma^{\EE}_{n\bk}(\varepsilon_{n\bk})$
along K$\to$M decreases with increasing $|\varepsilon_{n\bk}|$,
and it almost vanishes at the M point.
This strong $\bk$ anisotropy in the \EE\ contribution to the
imaginary part of the self-energy is a band structure effect, and is
absent in calculations based on the massless Dirac equation.
This behavior is in contrast with
the wavevector insensitivity of the phonon-induced electron
self-energy~\cite{park:2008PRB_Graphene_ElPh} (Fig.~\ref{Fig2}).
The calculated real part~\cite{rubio:037601} and the imaginary part~\cite{spataru:246405}
of the electron self-energy in bulk graphite arising from the \EE\ interaction
are also anisotropic, in line with the present findings.

Figures~\ref{Fig2}(a) and~\ref{Fig2}(b) show the electron self-energy
in {\it n}-doped graphene ($E_{\rm D}=-1.0$~eV) on a substrate
(model with $\varepsilon_{\rm b}=3.8$)
arising both from the \EE\ and the \EP\ interaction.
The Im~$\Sigma_{n\bk}(\varepsilon_{n\bk})$ along the two
different directions K$\Gamma$ and KM are qualitatively different
at high binding energy.
This anisotropy is due to the \EE\ interaction,
and not the \EP\ interaction~\cite{park:2008PRB_Graphene_ElPh}.
It is noted that the total linewidth along the KM direction
is almost constant for binding energies in the range 1.7 to 3.5~eV.
These anisotropic features should be observable in photoemission experiments.

The \EE\ and the \EP\ interactions give comparable
contributions to the imaginary part of the electron self-energy,
especially within a few tenths
of an eV from the Fermi level (Fig.~\ref{Fig2}).
This behavior is peculiar to graphene. In most metals
the \EP\ contribution to the electron self-energy near $E_{\rm F}$ is generally
dominant over the \EE\ contribution
at energies comparable to the relevant phonon
energy scale~\cite{note:elel_elph}.
Similarly large Im~$\Sigma_{n\bk}(\varepsilon_{n\bk})$
due to \EE\ interactions are obtained in the
Dirac Hamiltonian calculations in Refs.~\onlinecite{polini:081411}
and~\onlinecite{hwang:081412}
if the same background dielectric constant $\varepsilon_{\rm b}$
is used.
Because of this peculiar aspects of graphene, an \EP\ coupling strength $\lambda$
extracted from measured data could be overestimated
if the \EE\ interaction is neglected.
This may explain why the \EP\ coupling
strength $\lambda$
extracted from photoemission spectra~\cite{mcchesney:2007arxiv_Graphene_Anisotropy}
is larger
than the theoretical calculations~\cite{park:2008PRB_Graphene_ElPh,calandra:205411},
together with the effects of bare band curvature~\cite{park:2008PRB_Graphene_ElPh}
and dopants.

\begin{figure}
\includegraphics[width=1.0\columnwidth]{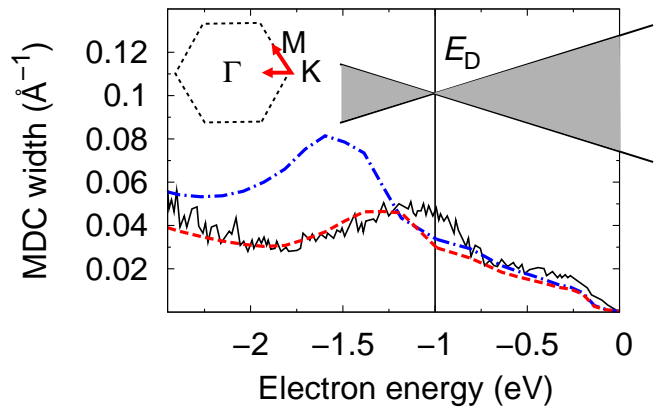}
\caption{(color online). MDC width
  versus binding energy in {\it n}-doped graphene ($E_{\rm D}=-1.0$~eV).
  Calculated quantities for suspended graphene ($\varepsilon_{\rm b}=1.0$) and
  for graphene on a model substrate ($\varepsilon_{\rm b}=3.8$) are shown in
  dash-dotted blue and dashed red lines, respectively.
  The experimental result
  measured for sample corresponding to the
  highest level of doping in Fig.~3 of
  Ref.~\onlinecite{bostwick:2007NatPhys}
  are shown as the solid black line~\cite{note:experiment_theory2}.
  Both the experimental and the calculated results are along the KM
  and the K$\Gamma$ direction of the Brillouin zone when the electron energy is
  above and below $E_{\rm D}$, respectively.}
\label{Fig3}
\end{figure}

We now compare the imaginary part of the electron self-energy obtained from
our calculation with the MDC width obtained from measured photoemission
spectra~\cite{bostwick:2007NatPhys}.
For a linear bare band energy dispersion,
the spectral function at a fixed energy $\omega$ is a Lorentzian as a function
of the wavevector measured from the $\bK$ point~\cite{bostwick:2007NatPhys}.
Thus, the width of the MDC $\Delta k$ at energy $\omega=\varepsilon_{n\bk}$
can be identified as
$\Delta k(\varepsilon_{n\bk})=2{\rm Im}\,\Sigma_{n\bk}(\varepsilon_{n\bk})/\hbar v_0$
where $v_0$ is the LDA band velocity of low-energy charge carriers in
graphene~\cite{bostwick:2007NatPhys,park:2007PRL_Graphene_ElPh}.
(For the {\it n}-doped graphene with
$E_{\rm D}=-1.0$~eV, the bare band dispersion is, to a good approximation, linear
in the energy range considered in Fig.~\ref{Fig3}.)

Figure~\ref{Fig3} shows the calculated MDC width for suspended graphene ($\varepsilon_{\rm b}=1.0$)
and for our model of graphene on SiC ($\varepsilon_{\rm b}=3.8$).
The substrate screening affects the position and the strength of the
peak arising from the electron-plasmon interaction,
while the low-energy part is insensitive to the dielectric screening from the substrate.
The calculated MDC width for graphene when substrate screening is accounted for is in
agreement with the experimental data of Ref.~\onlinecite{bostwick:2007NatPhys}
throughout the whole energy window shown in Fig.~\ref{Fig3}.
However, the experimentally measured MDC width in a 0.4~eV
energy window around $E_{\rm D}$ (=-1.0~eV)
is larger than that from our calculation.
This enhanced linewidth may possibly arise from
the gap which opens up at $E_{\rm D}$ and midgap states originating from the
interactions between graphene and
SiC substrate with a carbon buffer
layer~\cite{zhou:2007NatMat,mattausch:2007prl,varchon:126805,kim:176802}.

In conclusion, we have studied the electron linewidths of {\it n}-doped graphene
including both the \EE\ and the \EP\ interaction contributions, using
first-principles calculations.
The imaginary part of the electron self-energy arising from
the \EE\ interaction is strongly anisotropic in {\bf k}-space.
We have shown that for graphene,
unlike in conventional metals,
the \EE\ contribution is comparable
to the \EP\ contribution at low binding-energy.
Our calculation explains most of the scattering rate observed in
a recent photoemission experiment~\cite{bostwick:2007NatPhys};
however, near the Dirac point energy,
the calculated scattering rate is smaller than the measured one,
suggesting the possibility of band gap opening and midgap states.
These results contribute to the resolution of
the important controversy introduced earlier in this paper
and encourages further theoretical studies including both many-body interactions
and substrate effects at an atomistic level.
More generally, our first-principles calculations convincingly demonstrate 
that multiple many-body interactions 
ought to be considered on the same footing in order to achieve
a quantitative and comprehensive interpretation of high-resolution
angle-resolved photoemission spectra.

We thank A. Lanzara, J. L. McChesney, A. Bostwick,
T. Ohta, E. Rotenberg, S. Ismail-Beigi, E. H. Hwang and
J. D. Sau for fruitful discussions.
This work was supported by NSF Grant
No. DMR07-05941 and by the Director, Office of Science, Office of Basic Energy
Sciences, Division of Materials Sciences and Engineering Division,
U.S. Department of Energy under Contract No. DE-AC02-05CH11231.
Sandia is a multiprogram laboratory operated by Sandia Corporation,
a Lockheed Martin Company, for the United States Department of
Energy under Contract No. DE-AC01-94-AL85000.
Computational resources have been provided by NPACI and NERSC.
The calculations were performed using the
{\tt PARATEC}~\cite{paratec}, {\tt BerkeleyGW}~\cite{BerkeleyGW},
{\tt Quantum-Espresso}~\cite{baroni:2006_Espresso}
and {\tt Wannier}~\cite{mostofi:2006_Wannier} packages.


\begin{thebibliography}{10}
\bibitem{note:experiment_theory2}
We used the scaling relations given in
  Refs.~\onlinecite{polini:081411,hwang:081412,hwang:115434} to account for the
  difference between the experimental doping level ($E_{\rm D}=-0.87$~eV) and
  the one considered here ($E_{\rm D}=-1.0$~eV).
\end{thebibliography}
\end{document}